%% file: EXO-12-002_temp.tex
\pdfoutput=1

\documentclass[11pt,twoside,a4paper,cmspaper,final,collab]{cms-tdr}
\def\svnVersion{175953}\def\cmsCernNo{2012-299}\def\cmsCernDate{\today}\def\cmsMessage{Submitted to Physical Review Letters}

\begin{document}\cmsNoteHeader{EXO-12-002}

\hyphenation{had-ron-i-za-tion}
\hyphenation{cal-or-i-me-ter}
\hyphenation{de-vices}

\RCS$Revision: 165667 $
\RCS$HeadURL: svn+ssh://svn.cern.ch/reps/tdr2/papers/EXO-12-002/trunk/EXO-12-002.tex $
\RCS$Id: EXO-12-002.tex 165667 2013-01-19 19:11:24Z kkaadze $
\newlength\cmsFigWidth
\ifthenelse{\boolean{cms@external}}{\setlength\cmsFigWidth{0.45\textwidth}}{\setlength\cmsFigWidth{0.65\textwidth}}
\ifthenelse{\boolean{cms@external}}{\providecommand{\cmsLeft}{Top}}{\providecommand{\cmsLeft}{Left}}
\ifthenelse{\boolean{cms@external}}{\providecommand{\cmsRight}{Bottom}}{\providecommand{\cmsRight}{Right}}
\ifthenelse{\boolean{cms@external}}{%
\newcommand{\scotchrule[1]}{\centering\begin{ruledtabular}\begin{tabular}{#1}}
\newcommand{\donescotchrule}{\end{tabular}\end{ruledtabular}}
}{
\newcommand{\scotchrule[1]}{\centering\begin{tabular}{#1}\hline\hline}
\newcommand{\donescotchrule}{\hline\hline\end{tabular}}
}
\newcommand\ST{\ensuremath{S_\mathrm{{T}}}\xspace}

\cmsNoteHeader{EXO-12-002} 
\title{Search for pair production of third-generation leptoquarks and top squarks in \texorpdfstring{$\Pp\Pp$ collisions at $\sqrt{s}=7$\TeV}{pp collisions at sqrt(s) = 7 TeV}}

\date{\today}

\abstract{
Results are presented from a search for the pair production
of third-generation scalar and vector leptoquarks, as well as for top
squarks in R-parity-violating supersymmetric models. In either
scenario, the new, heavy particle decays into a
$\tau$ lepton and a \cPqb\ quark.
The search is based on a data sample of $\Pp\Pp$
collisions at $\sqrt{s}=7$\TeV,
which is collected by the CMS detector at the LHC and corresponds to
an integrated luminosity of 4.8\fbinv.
The number of observed events is found to be in agreement
with the standard model prediction, and exclusion limits on mass
parameters are obtained at the 95\% confidence level.
Vector leptoquarks with masses below 760\GeV are excluded and,
if the branching fraction of the scalar leptoquark decay
to $\tau$ lepton and \cPqb\ quark is assumed to be unity,
third-generation scalar leptoquarks with masses below 525\GeV are ruled out.
Top squarks with masses below 453\GeV are excluded
for a typical benchmark scenario, and limits on the
coupling between the top squark, $\tau$ lepton, and \cPqb\ quark,
$\lambda^{\prime}_{333}$ are obtained.
These results are the most stringent for these scenarios to date.
}

\hypersetup{%
pdfauthor={CMS Collaboration},
pdftitle={Search for pair production of third-generation leptoquarks and top squarks in pp collisions at sqrt(s) = 7 TeV},%
pdfsubject={CMS},%
pdfkeywords={CMS, physics, scalar, vector, leptoquarks, top squarks}}

\maketitle 

Many extensions~\cite{GUT,LQSU5,SU4,LQ3b,SUPERSTR,TC3}
of the standard model (SM)
predict new scalar or vector bosons, called leptoquarks,
which carry nonzero lepton and baryon numbers, as well
as color and fractional electric charge.
Such particles are motivated by a unified description of quarks and leptons.
The combination of both baryon and lepton numbers
implies that leptoquarks can mediate quark-lepton transitions,
and leptoquarks decay into a quark and a lepton
(with model-dependent branching fractions).
For leptoquark masses that are within reach of current
collider experiments, limits on flavor-changing neutral currents,
i.e. processes that change quark flavor
but not electric charge, along with limits
on other rare processes~\cite{LQ1},
favor leptoquarks that couple to quarks and
leptons within the same SM generation.

The dominant pair production mechanisms for leptoquarks
at the Large Hadron Collider (LHC) are gluon-gluon fusion
and quark-antiquark annihilation and the cross
sections for these processes depend only on the leptoquark
mass and spin.
The results are interpreted in
the context of models with either
scalar leptoquarks (LQ) or vector leptoquarks (VLQ).

Supersymmetry (SUSY)
is an attractive extension of the SM because it can resolve the
hierarchy problem~\cite{Martin97} without unnatural fine-tuning,
if the mass of the supersymmetric partner of the top quark (top squark,
or stop) is not too large~\cite{Papucci:2011wy}.
In this scenario, the large mixing angle between the
left-chiral and right-chiral stops ($\stL$ and $\stR$),
which arises from the
large top Yukawa coupling to the Higgs boson, can produce two mass
eigenstates, $\stone$ and $\sttwo$, with a large mass splitting.
Thus,
$M_{\stone}$ can be substantially smaller than the masses
of the other scalar SUSY particles.
This light-stop scenario can be realized in both R-parity-conserving (RPC)
and R-parity-violating (RPV) SUSY models,
where R-parity is a new, multiplicatively
conserved quantum number~\cite{Barbier20051}
that distinguishes SM and SUSY particles.
Most previous searches for the light stop have been performed in the
context of RPC models, in which the presence of two
undetected particles (the lightest supersymmetric particles)
generates a signature with large missing transverse momentum.
If R-parity is violated, however, supersymmetric particles can decay
into final states containing the standard model particles only.
These signatures are not considered in most
searches~\cite{CMS-SUSY1,CMS-SUSY2}.

At the LHC, a
${\stone}{\overline{\widetilde{\cPqt}}}_1$~pair is produced
via strong interactions. When the masses of the
supersymmetric partners of the gluon and quarks, excluding the top quark,
are large, the stop pair production cross section is
similar to that of the third-generation LQ.
The cross section also depends on the
first-generation squark mass and the stop mixing angle
because of loop corrections, but the contribution from these
diagrams is less than 2\%.
Trilinear RPV operators allow
the lepton-number-violating decay
$\stL\to\tau \cPqb$~\cite{Barbier20051} with a coupling
$\lambda^{\prime}_{333}$,
resulting in the same final state as for
third-generation LQ decay, with similar kinematics.

In this Letter, a search is presented for pair production
of third-generation leptoquarks or stops
each decaying to a $\tau$ lepton
and a \cPqb~quark, using $\Pp\Pp$ collision data at $\sqrt{s}=7$\TeV.
The data sample has been recorded by the CMS detector and corresponds
to an integrated luminosity of 4.8\fbinv.
One of the $\tau$ leptons in the final state is required
to decay leptonically, $\tau\to \ell\cPgn_{\ell} \cPgn_{\tau}$, where $\ell$
can be either a muon or an electron, referred to as the light lepton below.
The other $\tau$ lepton is required
to decay to hadrons ($\tau_{\mathrm{h}}$),
$\tau \to \text{hadrons} + \nu_{\tau}$.
These requirements result in two possible final states referred
to as $\Pe\tau_{\mathrm{h}} \bbbar$ and $\mu\tau_{\mathrm{h}} \bbbar$.
The experimental signature is characterized by an energetic electron or muon,
a $\tau_{\mathrm{h}}$, and two jets produced by the
hadronization of \cPqb\ quarks (\cPqb~jets).
For the pair production of leptoquarks or stops, the scalar sum of the
transverse momenta (\pt) of the decay products,
$\ST\equiv \pt^{\tau_{\mathrm{h}}}+\pt^{\ell}+\pt^{\cPqb_1}+\pt^{\cPqb_2}$,
is expected to be large, as is the invariant mass of each system
containing a \cPqb\ jet and a $\tau$ lepton originating from the same heavy
particle.

No evidence for third-generation LQ or
stops has been found in previous searches,
using a final state with $\tau_{\mathrm{h}}$,
light lepton, and two \cPqb~jets.
The most stringent lower limits on
LQ and stop masses are 210\GeV~\cite{D0LQ3a}
and 153\GeV~\cite{CDF-Stop}, respectively.
A search performed by the CMS Collaboration
has excluded the existence of a third-generation LQ
with an electric charge of $\pm$1/3 and mass below 450\GeV,
assuming 100\% branching fraction to a \cPqb\ quark and a
$\nu_{\tau}$~\cite{CMS-PAS-EXO-11-030}.
Indirect bounds~\cite{lambdalimit}
exclude the region $\lambda^{\prime}_{333} > 0.26$
for $M_{\stone}\sim100$\GeV.

The central feature of the CMS
apparatus is a superconducting solenoid, of 6\unit{m} internal diameter, providing
a field of 3.8\unit{T}. A silicon pixel and strip tracker,
which allows the reconstruction of the trajectories
of charged particles within the pseudorapidity range $|\eta|<2.5$,
where
$\eta = -\ln [\tan(\theta/2)]$ and $\theta$ is the polar angle
with respect to the counterclockwise proton beam, are the innermost
parts of the CMS detector. The tracker is surrounded by
a calorimetry system, consisting of
a lead-tungstate crystal electromagnetic
calorimeter (ECAL) and  a brass/scintillator hadron calorimeter,
which measures particle energy
depositions for $|\eta| < 3$. The tracker and ECAL are placed within the
superconducting solenoid.
Muons are identified in gas-ionization
detectors embedded in the steel flux return yoke of the magnet.
Collision events are selected
using a two-tiered trigger system.
A more detailed description
of the CMS detector can be found in Ref.~\cite{CMSdetector}.

Events are collected using triggers requiring the presence
of an electron or a muon
and a $\tau_{\mathrm{h}}$ with transverse momentum thresholds ranging between
12--20\GeV
and 15--20\GeV, respectively, depending on the data-taking period.
Electrons are reconstructed using the tracker and
fully instrumented barrel ($|\eta|<1.44)$) or
endcap ($1.57<|\eta|<2.1$) regions of the ECAL.
Selected electrons are required to have
transverse momenta $\pt > 30$\GeV,
an electromagnetic shower shape consistent with
that of an electron, and an ECAL energy deposition that is compatible
with the track reconstructed in the tracker.
Muons are required to be reconstructed
by both the tracker and the muon spectrometer.
Candidates are required to have
$|\eta|<2.1$ and $\pt > 30$\GeV.
A particle-flow (PF)
technique~\cite{CMS-PAS-PFT-09-001}
is used for the reconstruction of $\tau_{\mathrm{h}}$
candidates. Information from all subdetectors
is combined to reconstruct
and identify final-state particles (PF candidates)
produced in the collision.
The PF candidates are used with
the hadron-plus-strips algorithm~\cite{HPS} to reconstruct
hadronic decays of $\tau$ leptons with one or
three charged pions and up to two neutral pions.
The reconstructed $\tau_{\mathrm{h}}$
is required to have $\pt> 50$\GeV and
$|\eta|<2.3$. The light lepton and $\tau_{\mathrm{h}}$ are required
to have opposite electric charge.
To reduce background from additional proton-proton interactions
in the same beam crossing (pileup),
the light lepton and $\tau_{\mathrm{h}}$ are required to originate from
the same vertex. The criteria for association
to the vertex are optimized to take into account the
finite lifetime of the $\tau$ lepton and are efficient
for selecting an electron or muon from its decay.
Selected electrons,
muons, and $\tau_{\mathrm{h}}$ are required to
be isolated from other PF candidates and to be separated
by $\Delta R \equiv \sqrt{(\Delta\phi)^2 + (\Delta\eta)^2} > 0.5$
for both the $\Pe\tau_{\mathrm{h}} \bbbar$ and $\mu\tau_{\mathrm{h}} \bbbar$ channels.
Here, $\Delta\phi$ is an azimuthal and $\Delta\eta$ is a
pseudorapidity separation between the light lepton and $\tau_{\mathrm{h}}$.

Jets are reconstructed using PF candidates with the
anti-$kT$ algorithm~\cite{Cacciari:2008gp} with a
distance parameter of 0.5.
An average contribution of pileup interactions
is estimated and subsequently subtracted from the jet
energy~\cite{Cacciari:JetArea}.
Selected jets are required to be within $|\eta|<2.4$ and
have $\pt>30$\GeV. Additionally, these jets must be
separated from the selected light lepton and $\tau_{\mathrm{h}}$ by
$\Delta R>0.5$.
The selected events are required to have at least two jets identified as
originating from \cPqb\ quark hadronization (\cPqb-tagged) using a
displaced track counting algorithm,
based on track impact parameter significance~\cite{BTV-11-004}.

To discriminate between signal and background,
the invariant mass of the $\tau_{\mathrm{h}}$ and \cPqb~jet
($M_{\tau_{\mathrm{h}},\cPqb}$)
is required to be greater than 170\GeV.
Of the two possible pairings of the $\tau_{\mathrm{h}}$ and \cPqb~jets,
the one for which
the invariant mass is closest to the invariant mass of the
light lepton and
the other \cPqb~jet is chosen as an observable.
After the final selection, the \ST distribution is used to search for
an excess above the SM expectation.

The dominant sources of $\ell\tau_{\mathrm{h}} \bbbar$ events from
SM processes are
the production of a \PW{} or \cPZ{} boson associated with jets,
where a jet is misidentified as a $\tau_{\mathrm{h}}$, and
\ttbar pair production.
There is also a small contributions from
\cPZ{} bosons decaying to a pair of $\tau$ leptons,
or to a pair of electrons or muons,
where one of the electrons or muons is misidentified as the $\tau_{\mathrm{h}}$,
and from single-top and diboson production processes.

The LQ signal is generated using the \PYTHIA
(v6.420)~\cite{Sjostrand:2003wg}
generator for a range of
leptoquark masses $M_{LQ}$ spanning 150 to 800\GeV.
The \MADGRAPH generator~\cite{Alwall:2011uj} interfaced with
\PYTHIA for hadronization and showering is used to model the dominant
$\ttbar$ and \PW+jets backgrounds. These generators are also used
to model the less significant Drell--Yan process $\cPZ/\gamma^*$+jets.
The single top production is modeled with the \POWHEG~\cite{POWHEG2}
generator, and diboson processes are modeled with
\PYTHIA v6.4. All generated samples are interfaced with
\TAUOLA~\cite{TAUOLA} for $\tau$ decay, passed through a full detector
simulation based on \GEANTfour~\cite{Agostinelli2003250} and the complete
reconstruction chain used for data analysis.
The VLQ and stop pair production processes are modeled using
the \textsc{CalcHEP}~\cite{LQVLQ,CalcHEP}
and \textsc{prospino}~\cite{prospino}
generators in order to compare the kinematics of their decay products
with those from scalar LQ.
The most precise available cross section calculations,
either next-to-leading order (NLO) or next-to-NLO, are used
to normalize the signal~\cite{LQ-sigmaNLO}
and background processes~\cite{FEWZ,Kidonakis:2010}.

The efficiencies of the trigger and final selection criteria
for signal processes are
estimated from the simulation.
The identification efficiencies for leptons and
\cPqb~jets are found from data in different data-taking periods,
and used where necessary to correct the event selection efficiency
estimates from the simulation.
The trigger efficiency for signal events with a LQ mass hypothesis of 550\GeV
is close to 90\% for both
channels. The efficiency of the final selection is
$(8.4\pm0.2\stat\pm0.6\syst)\%$ and
$(13.3\pm0.3\stat\pm0.9\syst)\%$ for the
$\Pe\tau_{\mathrm{h}} \bbbar$~and~$\mu\tau_{\mathrm{h}} \bbbar$ channels, respectively.

The \ttbar background is estimated using simulation.
The normalization and several kinematic distributions of the
\ttbar background are validated using
events rejected by the $M_{\tau_{\mathrm{h}},\cPqb}>170$\GeV
criterion.
Both the yield and the \ST distribution in this control region agree well
with the data observation.

The number of \PW{} or \cPZ{} background events containing a jet
misidentified as a $\tau_{\mathrm{h}}$ is estimated from data.
The probability of misidentification is measured
using events with a \PW\ boson produced in association with one jet
passing $\tau_{\mathrm{h}}$ selection criteria except the
isolation requirement.
The decay to electron or muon of the \PW\ boson is used.
In the selected sample, the lepton is required to be well identified.
The transverse mass ${M_T = (2p_{\mathrm{T}}^{\ell}{\ETslash}(1-\cos(\Delta\phi)))^{1/2}}$
is required to be greater than 50\GeV.
Here, $\pt^{\ell}$ and
$\ETslash$
are the transverse momentum of the lepton,
and the imbalance of the transverse energy in the event, respectively,
and $\Delta\phi$ is the azimuthal angle between the lepton and the
$\ETslash$
direction.
To reduce the contribution
from \ttbar events, the candidate $\tau_{\mathrm{h}}$ and the lepton
are required to
have the same electric charge.
The probability to satisfy the final $\tau_{\mathrm{h}}$ selection
criteria is found to be independent of the
transverse momentum and pseudorapidity
of the candidate $\tau_{\mathrm{h}}$, and is  $f = (2.44 \pm 0.53)\%$.
The number of background events is given by $N_{\text{bkg}} = N_{\PW_{\tau_{\mathrm{h}}}}\times f/(1-f)$, where
$N_{\PW_{\tau_{\mathrm{h}}}}$ is the number of events in the control
sample with well identified lepton, two \cPqb~jets, and
a $\tau$ candidate that passes
the $\tau_{\mathrm{h}}$ identification criteria but fails the
isolation requirement
and has opposite electric charge
to that of the lepton.
The contribution from \ttbar background in this sample is subtracted.
The \ST distribution for this background is determined
from the Monte Carlo (MC) simulation. Because of statistical limitations
on the MC samples,
events with a lepton, a $\tau_{\mathrm{h}}$ candidate, and two jets are used,
and the jet \pt spectrum is reweighted to match that
expected from \cPqb~jets.

The small background processes, such as $\cPZ\to\tau\tau$
and dibosons decaying into genuine $\tau_{\mathrm{h}}$
or $\cPZ\to\Pe\Pe$ and $\cPZ\to\mu\mu$,
with a light lepton misidentified as a $\tau_{\mathrm{h}}$,
are estimated using the simulated data.

The estimations of the background and the signal efficiency
are affected by systematic uncertainties.
The uncertainty in the total integrated luminosity is
2.2\%~\cite{CMS-PAS-SMP-12-008}.
The uncertainty in the
trigger and lepton efficiencies
is 1--3\%. The uncertainty assigned to the $\tau_{\mathrm{h}}$
identification efficiency is 6\%,
while the
uncertainty in the b-tagging efficiency and mistagging probability are 4\% and 10\%,
respectively.
Systematic uncertainties of 17\% and 13\% are assigned
to the normalization of the \ttbar
background in the ${\Pe}\tau_{\mathrm{h}}\bbbar$
and the $\mu\tau_{\mathrm{h}}\bbbar$ channels, respectively,
based on the statistical uncertainties of the control sample
and the uncertainties in the MC prediction,
to which it is compared.
The uncertainty in the cross section measurements
for diboson production~\cite{CMS-EWK-DIW}
is 30\%, leading to a normalization uncertainty in the
corresponding background rate.
Owing to the statistical limitation on
$\cPZ\to\tau\tau/\ell\ell$ simulation, the uncertainty
in these backgrounds is 70\% and 30\%
for the $\Pe\tau_{\mathrm{h}}\cPqb\cPqb$ 
and $\mu\tau_{\mathrm{h}}\cPqb\cPqb$,
respectively. A 40\% systematic uncertainty is assigned to the
modeling of Z production
in association with two \cPqb\ jets~\cite{CMS-EWK-11-012}.
A 4\% uncertainty,
due to modeling of initial- and final-state radiation
in the simulation,
is assigned to the signal acceptance. Uncertainty due to
the effect of pileup modeling in the MC simulations is
estimated to be 3\%.
Jet energy scale (2--4\%
depending on pseudorapidity and transverse momentum)
as well as energy scale (3\%) and resolution (10\%) uncertainties
for $\tau_{\mathrm{h}}$ which affect both the \ST distribution
and the expected yields
from the signal and background processes are taken into account.

Uncertainties due to the choice of parton distribution
functions (PDF) of the proton lead to changes in the
total cross section and the acceptance for both
signal and background processes.
PDF uncertainties in the theoretical cross section and
on the final-state acceptance are calculated using
the PDF4LHC~\cite{PDF4LHC1} prescription, and
are found to vary between 10--30\% and 1--3\%, respectively.

\begin{figure}[htb]
  \centering
    \includegraphics[width=\cmsFigWidth]{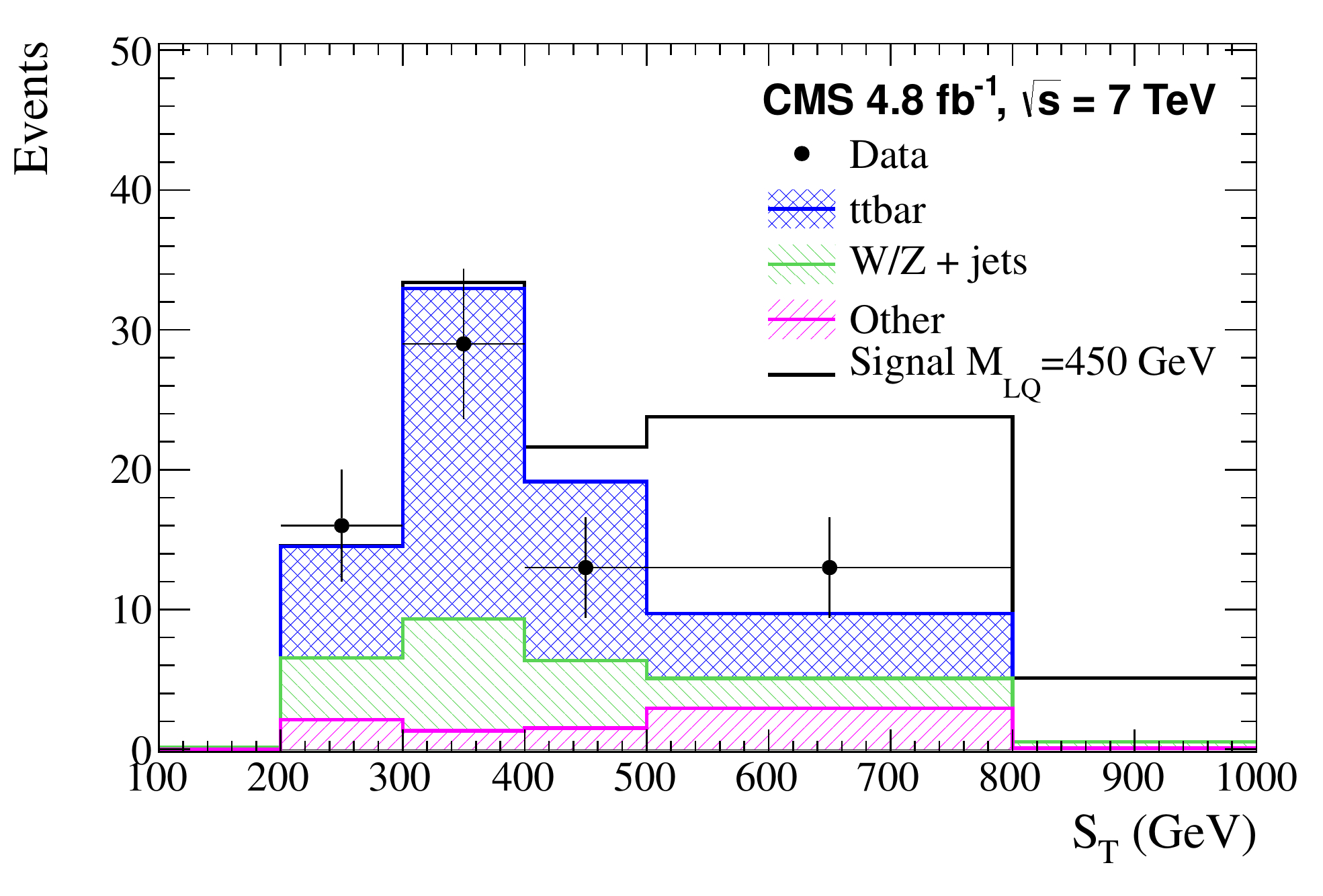}
    \caption{The measured \ST distribution (points) compared to the stacked distribution of the SM backgrounds (shaded region) and a simulated  $M_{\mathrm{LQ}}=450$\GeV LQ signal (solid line) after the final selection (color online).}
    \label{fig:STfinal}
\end{figure}

\begin{figure}[htb]
  \centering
  \includegraphics[width=0.4\textwidth]{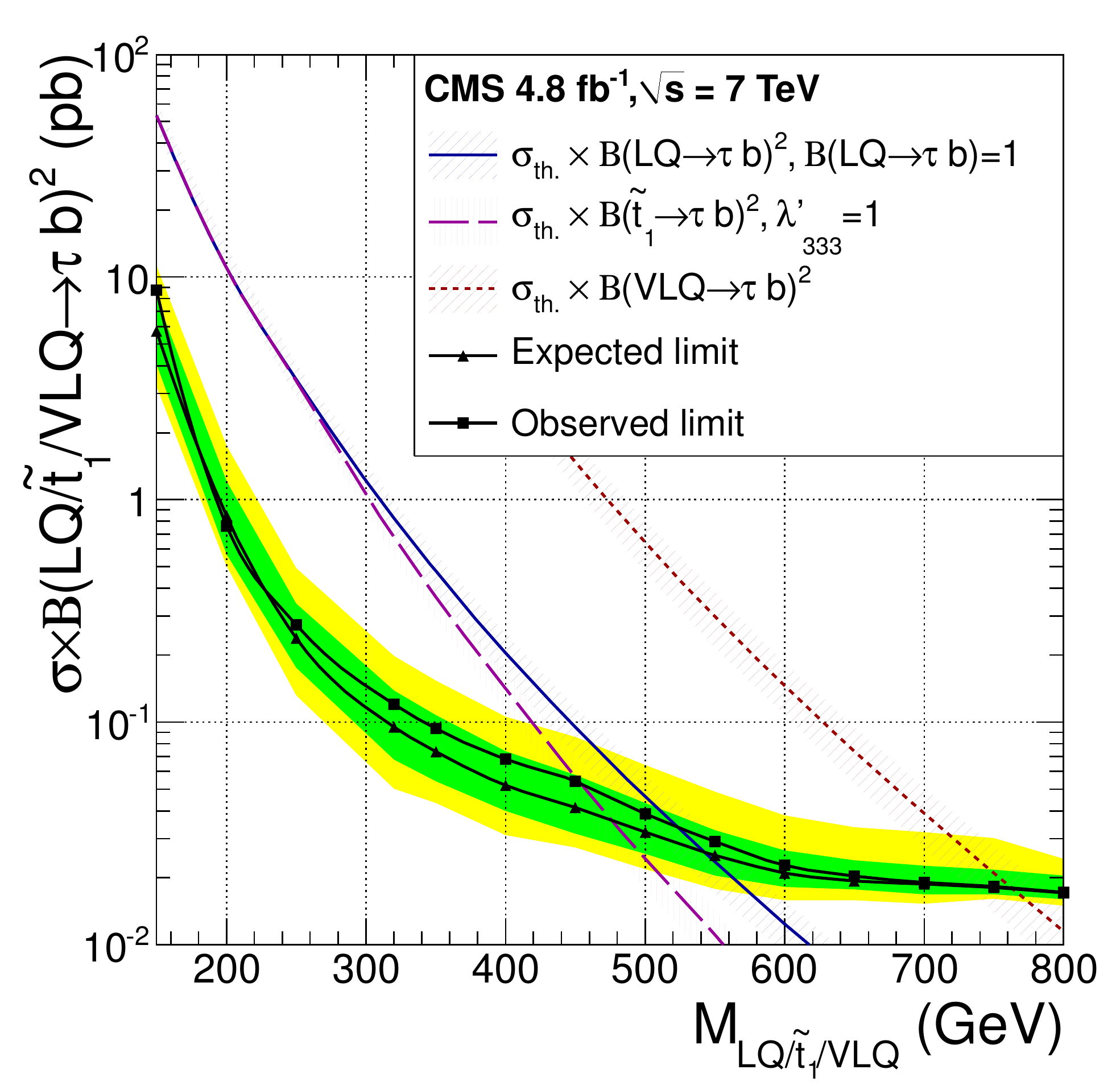}
  \includegraphics[width=0.4\textwidth]{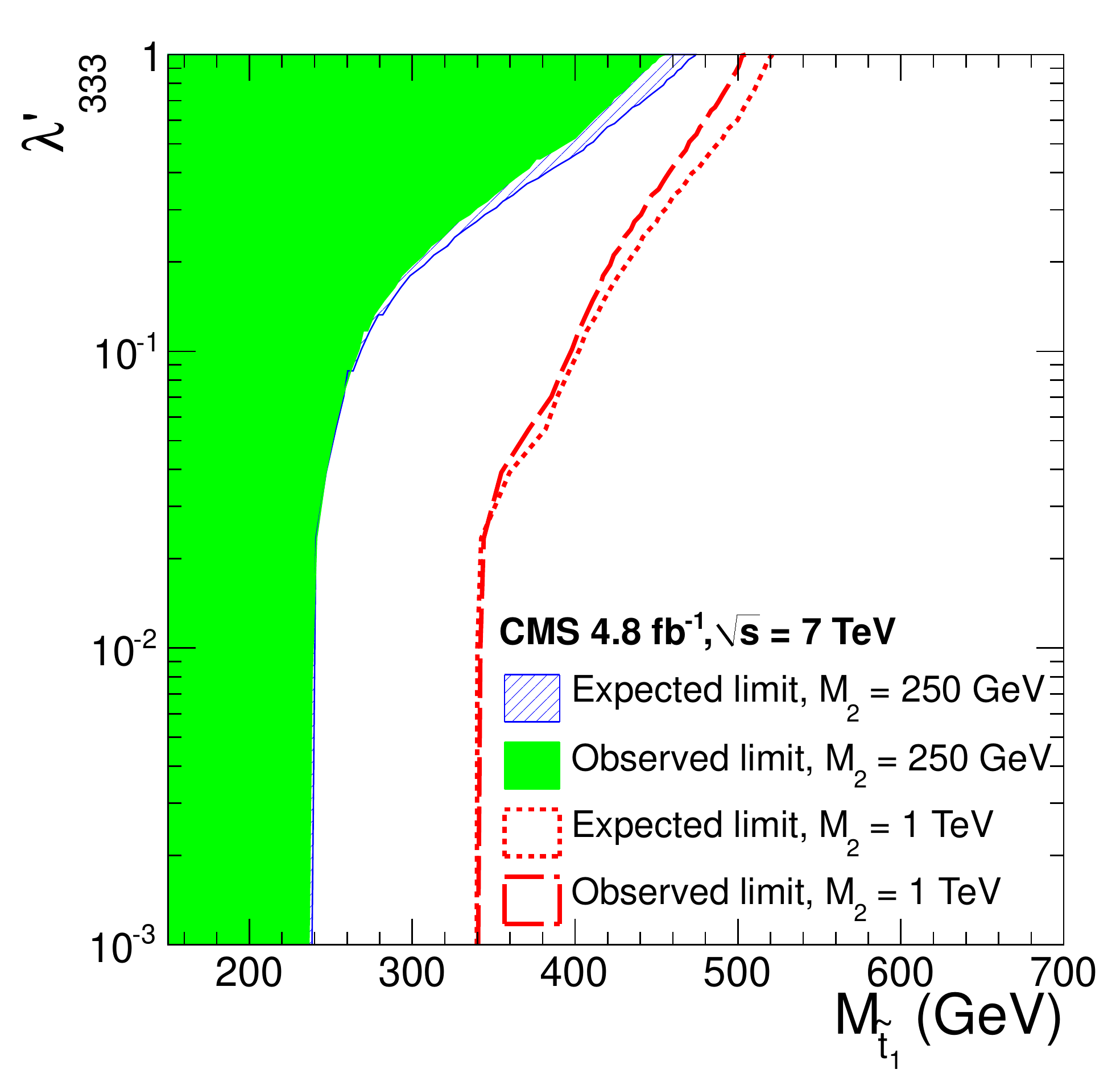}
\caption{\cmsLeft: the expected and observed upper limit at 95\% CL on the LQ ($\stone$, VLQ) pair production cross section times $\mathcal{B}(\mathrm{LQ}/\stone/\mathrm{VLQ}\to \tau \cPqb)$ as a function of the LQ ($\stone$, VLQ) mass. The $\pm1\,\sigma$ and $\pm2\,\sigma$ uncertainties on the expected limit are also shown as green (inner) and yellow (outer) bands around the expected limit. The blue (solid) curve, magenta (dashed) curve, and red (dotted) curve and the matching shaded bands represent the theoretical LQ, $\stone$, and VLQ pair production cross section and the uncertainties due to the choice of PDF and renormalization and factorization scales, respectively. \cmsRight: the expected and observed 95\% CL limit on the RPV coupling $\lambda^{\prime}_{333}$ for $M_2$ = 250\GeV and $M_2 = 1$\TeV (color online).}
    \label{fig:comblimitAsymptotic}
\end{figure}

The number of observed events and the expected signal and
background yields after the final selection
are listed in Table~\ref{tab:FinalYields170}.
Data are in good agreement with the SM background prediction.
The \ST distribution of selected events in data and MC simulation
is shown in Fig.~\ref{fig:STfinal}.
As the distribution of \ST predicted for the SM background is in good agreement
with the distribution obtained in data, a limit is set on the
product of the cross section for pair production of
third-generation LQ and the square of the
branching fraction for the decay to $\tau$ lepton and \cPqb\ quark.
The modified frequentist construction
CL$_\mathrm{s}$~\cite{LHC-HCG}
is used for limit calculation.
A maximum likelihood fit is performed to
the \ST spectrum simultaneously for
both $\Pe\tau_{\mathrm{h}} \bbbar$ and
$\mu\tau_{\mathrm{h}} \bbbar$ channels, taking into account
correlations between the systematic uncertainties.
The limits as a function of the LQ mass
are shown in Fig.~\ref{fig:comblimitAsymptotic}.
Assuming  $\mathcal{B}({\rm LQ}~\to\tau\rm{b})=1$,
we exclude LQ
with masses below 525\GeV at 95\% Confidence Level (CL),
in good agreement with the expected limit at 543\GeV.
The difference between acceptance and selection efficiency for
LQ and VLQ is less than a few percent~\cite{Brooijmans:2012yi}.
Thus, the same observed limit can be
used to extract the limit on a top SU(5)
VLQ predicted by the model of Ref.~\cite{LQSU5}.
Such vector leptoquarks with masses 760\GeV are excluded
at 95\% CL, in agreement with the expected limit of 762\GeV.

\begin{table}[!ht]
    \caption{Estimated signal (LQ) and background yields and observed events in data after the final selection. The first value in the uncertainty on the yield is the statistical contribution and the second value is the systematic contribution. The PDF uncertainties are not included.}
\renewcommand{\arraystretch}{1.2}
    \scotchrule[lcc]
      & $\mu+\tau_{\mathrm{h}}\bbbar$ channel & $\Pe+\tau_{\mathrm{h}}\bbbar$ channel \\
      \hline

      \ttbar          & $38.1\pm3.4\pm5.7$  & $10.9\pm1.8\pm2.0$\\
      W+jets/Z+jets   & $11.6\pm0.1\pm3.6$  & $8.4\pm0.1\pm2.6$ \\
      Z($\tau\tau$/$ll$)& $5.0\pm1.6\pm2.1$ & $2.1\pm1.5\pm0.9$ \\
      Diboson         & $0.5\pm0.1\pm0.2$   & $0.3\pm0.1\pm0.1$ \\
      \hline
      Total background & $55.2\pm3.8\pm7.5$ & $21.8\pm2.3\pm3.6$ \\
      \hline
      \hline
      Data          & $46$ & $25$ \\
      \hline
      \hline
      Signal (450\GeVns) & $13.2\pm0.3\pm0.9$ & $8.4\pm0.2\pm0.6$ \\
  \donescotchrule
\renewcommand{\arraystretch}{1.}
    \label{tab:FinalYields170}
\end{table}

These results are interpreted as a limit on stop pair production
with RPV decay.
Assuming $\mathcal{B}(\stone\rightarrow\tau\cPqb) = 1$, stop masses below
525\GeV are excluded.
A limit is also extracted for a benchmark scenario,
where the branching ratio
$\mathcal{B}(\stone\rightarrow\tau\cPqb)$ decreases as stop
mass increases as
R-parity-conserving decays open up. The MSSM parameters used in a
benchmark scenario are
heavy SU(2) gaugino $M_2 = 250$\GeV,
heavy Higgsino mixing parameter $\mu = 380$\GeV, tan$\beta = 40$,
where $\beta$ is the ratio of the Higgs vacuum
expectation values, stop mixing angle $\theta=0$, and
$\lambda^{\prime}_{333} = 1$.
The limit on
$\sigma\mathcal{B}(\stone\rightarrow\tau\cPqb)^2$
as a function of stop mass is shown in
Fig.~\ref{fig:comblimitAsymptotic}.
Using this benchmark, the R-parity-violating stop is
excluded for masses below 453\GeV in agreement with an
the expected exclusion mass of 474\GeV.
Using the same parameter set, but two $M_2$ values
(250\GeV and 1\TeV), limits are set on RPV coupling
$\lambda^{\prime}_{333}$ as a function of stop mass.
The results are shown in Fig~\ref{fig:comblimitAsymptotic}.
Top squarks with mass below 240\GeV (340\GeV)
are excluded for $M_2$ = 250\GeV ($M_2 = 1$\TeV)
for all values of
$\lambda^{\prime}_{333} > \mathcal{O}(10^{-7})$,
corresponding to a decay length of about 0.5\mm.
Stops with very small values of $\lambda^{\prime}_{333}$ have been
excluded by a different CMS
analysis~\cite{springerlink:10.1007/JHEP08(2012)026}.

In summary, a search for pair production of third-generation
scalar and vector leptoquarks and top squarks decaying in a
RPV scenario has been presented. The search is
performed in the final state including an electron or a muon, a
hadronically decaying $\tau$ lepton and two \cPqb\ jets.
No excess above the
SM background prediction is observed at high \ST.
Assuming a 100\% branching fraction
to a $\tau$ lepton and a \cPqb\ quark, the existence of the scalar leptoquarks
with masses below 525\GeV is excluded at 95\% CL.
The existence of SU(5) vector leptoquarks with masses below 760\GeV is also
excluded at 95\% CL.
Limits are also set on top squark pair production with RPV decay.
The limits are obtained on $\lambda^\prime_{333}$ as a function of stop mass,
and stops with masses below 453\GeV are excluded for a benchmark scenario
with $\lambda^\prime_{333}=1$.
These limits are the most stringent to date, and the limits
on $\lambda^\prime_{333}$ are the first direct limits that
significantly improve previous indirect bounds.

We wish to congratulate our colleagues in the CERN accelerator departments
for the excellent performance of the LHC machine. We thank the technical
and administrative staffs at CERN and other CMS institutes, and acknowledge
support from: FMSR (Austria); FNRS and FWO (Belgium); CNPq, CAPES, FAPERJ,
and FAPESP (Brazil); MES (Bulgaria); CERN; CAS, MoST, and NSFC (China);
COLCIENCIAS (Colombia); MSES (Croatia); RPF (Cyprus); Academy of Sciences
and NICPB (Estonia); Academy of Finland, ME, and HIP (Finland);
CEA and CNRS/IN2P3 (France); BMBF, DFG, and HGF (Germany);
GSRT (Greece); OTKA and NKTH (Hungary); DAE and DST (India); IPM (Iran);
SFI (Ireland); INFN (Italy); NRF and WCU (Korea); LAS (Lithuania);
CINVESTAV, CONACYT, SEP, and UASLP-FAI (Mexico); PAEC (Pakistan);
SCSR (Poland); FCT (Portugal); JINR (Armenia, Belarus, Georgia, Ukraine,
Uzbekistan); MST and MAE (Russia); MSTD (Serbia); MICINN and CPAN (Spain);
Swiss Funding Agencies (Switzerland); NSC (Taipei); TUBITAK and
TAEK (Turkey); STFC (United Kingdom); DOE and NSF (USA).

\bibliography{auto_generated}

\cleardoublepage \appendix\section{The CMS Collaboration \label{app:collab}}\begin{sloppypar}\hyphenpenalty=5000\widowpenalty=500\clubpenalty=5000\input{EXO-12-002-authorlist.tex}\end{sloppypar}
\end{document}

%% file: EXO-12-002-authorlist.tex
\textbf{Yerevan Physics Institute,  Yerevan,  Armenia}\\*[0pt]
S.~Chatrchyan, V.~Khachatryan, A.M.~Sirunyan, A.~Tumasyan
\vskip\cmsinstskip
\textbf{Institut f\"{u}r Hochenergiephysik der OeAW,  Wien,  Austria}\\*[0pt]
W.~Adam, E.~Aguilo, T.~Bergauer, M.~Dragicevic, J.~Er\"{o}, C.~Fabjan\cmsAuthorMark{1}, M.~Friedl, R.~Fr\"{u}hwirth\cmsAuthorMark{1}, V.M.~Ghete, J.~Hammer, N.~H\"{o}rmann, J.~Hrubec, M.~Jeitler\cmsAuthorMark{1}, W.~Kiesenhofer, V.~Kn\"{u}nz, M.~Krammer\cmsAuthorMark{1}, I.~Kr\"{a}tschmer, D.~Liko, I.~Mikulec, M.~Pernicka$^{\textrm{\dag}}$, B.~Rahbaran, C.~Rohringer, H.~Rohringer, R.~Sch\"{o}fbeck, J.~Strauss, A.~Taurok, W.~Waltenberger, G.~Walzel, E.~Widl, C.-E.~Wulz\cmsAuthorMark{1}
% ...
\vskip\cmsinstskip
\dag:~Deceased\\
1:~~Also at Vienna University of Technology, Vienna, Austria\\
% ...


%% file: EXO-12-002_temp.bbl
\providecommand{\href}[2]{#2}\begingroup\raggedright\begin{thebibliography}{10}%
\makeatletter
\providecommand{\hrefCMSnoop }[0]{\@secondoftwo}%
\makeatother
\providecommand{\doi}{\texttt{doi:}\begingroup \urlstyle{tt}\Url}

\bibitem{GUT}
\hrefCMSnoop {} {H.~Georgi and S.~L. Glashow, ``Unity of All
  Elementary-Particle Forces'',} \textit{ Phys. Rev. Lett.} \textbf{ 32} (1974)
  438,
  \href{http://dx.doi.org/10.1103/PhysRevLett.32.438}{\doi{10.1103/PhysRevLett.32.438}}.

\bibitem{LQSU5}
S.~Chakdar\hrefCMSnoop {} { {et~al.}, ``Unity of elementary particles and
  forces for the third family'',} \textit{ Phys. Lett. B} \textbf{ 718} (2012)
  121,
  \href{http://dx.doi.org/10.1016/j.physletb.2012.10.021}{\doi{10.1016/j.physletb.2012.10.021}},
  \href{http://www.arXiv.org/abs/1206.0409}{\texttt{ arXiv:1206.0409}}.

\bibitem{SU4}
\hrefCMSnoop {} {J.~C. Pati and A.~Salam, ``Lepton number as the fourth
  `color''',} \textit{ Phys. Rev. D} \textbf{ 10} (1974) 275,
  \href{http://dx.doi.org/10.1103/PhysRevD.10.275}{\doi{10.1103/PhysRevD.10.275}}.

\bibitem{LQ3b}
\hrefCMSnoop {} {B.~Gripaios, ``Composite leptoquarks at the LHC'',} \textit{
  JHEP} \textbf{ 02} (2010) 045,
  \href{http://dx.doi.org/10.1007/JHEP02(2010)045}{\doi{10.1007/JHEP02(2010)045}}.

\bibitem{SUPERSTR}
\hrefCMSnoop {} {J.~L. Hewett and T.~G. Rizzo, ``{Low-energy phenomenology of
  superstring inspired E$_6$ models}'',} \textit{ Phys. Rept.} \textbf{ 183}
  (1989) 193,
\href{http://dx.doi.org/10.1016/0370-1573(89)90071-9}{\doi{10.1016/0370-1573(89)90071-9}}.

\bibitem{TC3}
\hrefCMSnoop {} {E.~Eichten and K.~Lane, ``Dynamical breaking of weak
  interaction symmetries'',} \textit{ Phys. Lett. B} \textbf{ 90} (1980) 125,
  \href{http://dx.doi.org/10.1016/0370-2693(80)90065-9}{\doi{10.1016/0370-2693(80)90065-9}}.

\bibitem{LQ1}
\hrefCMSnoop {} {O.~Shanker, ``{$\pi \ell2$, K$\ell3$~and
  $\rm{K^{0}}-\rm{\bar{K}^{0}}$ constraints on leptoquarks and supersymmetric
  particles}'',} \textit{ Nucl. Phys. B} \textbf{ 204} (1982) 375,
  \href{http://dx.doi.org/10.1016/0550-3213(82)90196-1}{\doi{10.1016/0550-3213(82)90196-1}}.

\bibitem{Martin97}
\hrefCMSnoop {} {S.~P. Martin, ``A Supersymmetry Primer'',} (1997).
  \href{http://www.arXiv.org/abs/hep-ph/9709356}{\texttt{
  arXiv:hep-ph/9709356}}. See also references therein.

\bibitem{Papucci:2011wy}
\hrefCMSnoop {} {M.~Papucci, J.~T. Ruderman, and A.~Weiler, ``{Natural SUSY
  Endures}'',} \textit{ JHEP} \textbf{ 09} (2012) 035,
  \href{http://dx.doi.org/10.1007/JHEP09(2012)035}{\doi{10.1007/JHEP09(2012)035}},
\href{http://www.arXiv.org/abs/1110.6926}{\texttt{ arXiv:1110.6926}}.

\bibitem{Barbier20051}
R.~Barbier\hrefCMSnoop {} { {et~al.}, ``R-Parity-violating supersymmetry'',}
  \textit{ Phys. Rept.} \textbf{ 420} (2005) 1,
  \href{http://dx.doi.org/10.1016/j.physrep.2005.08.006}{\doi{10.1016/j.physrep.2005.08.006}},
  \href{http://www.arXiv.org/abs/hep-ph/0406039}{\texttt{
  arXiv:hep-ph/0406039}}.

\bibitem{CMS-SUSY1}
\hrefCMSnoop {} {{ CMS} Collaboration, ``Search for supersymmetry in events
  with a lepton, a photon, and large missing transverse energy in pp collisions
  at $\sqrt{s} = 7$~TeV'',} \textit{ JHEP} \textbf{ 06} (2011) 093,
  \href{http://dx.doi.org/10.1007/JHEP06(2011)093}{\doi{10.1007/JHEP06(2011)093}},
  \href{http://www.arXiv.org/abs/1105.3152}{\texttt{ arXiv:1105.3152}}.

\bibitem{CMS-SUSY2}
\hrefCMSnoop {} {{ CMS} Collaboration, ``Search for Supersymmetry at the LHC in
  Events with Jets and Missing Transverse Energy'',} \textit{ Phys. Rev. Lett.}
  \textbf{ 107} (2011) 221804,
  \href{http://dx.doi.org/10.1103/PhysRevLett.107.221804}{\doi{10.1103/PhysRevLett.107.221804}},
  \href{http://www.arXiv.org/abs/1109.2352}{\texttt{ arXiv:1109.2352}}.

\bibitem{D0LQ3a}
\hrefCMSnoop {} {{ D0} Collaboration, ``{Search for Third Generation Scalar
  Leptoquarks Decaying into $\tau$b}'',} \textit{ Phys. Rev. Lett.} \textbf{
  101} (2008) 241802,
  \href{http://dx.doi.org/10.1103/PhysRevLett.101.241802}{\doi{10.1103/PhysRevLett.101.241802}},
  \href{http://www.arXiv.org/abs/0806.3527}{\texttt{ arXiv:0806.3527}}.

\bibitem{CDF-Stop}
\hrefCMSnoop {} {{ CDF} Collaboration, ``{Search for Pair Production of Scalar
  Top Quarks Decaying to a $\tau$ Lepton and a $b$ Quark in $p\overline{p}$
  Collisions at $\sqrt{s}=1.96\text{\,}\text{\,}\mathrm{TeV}$}'',} \textit{
  Phys. Rev. Lett.} \textbf{ 101} (2008) 071802,
  \href{http://dx.doi.org/10.1103/PhysRevLett.101.071802}{\doi{10.1103/PhysRevLett.101.071802}},
  \href{http://www.arXiv.org/abs/0802.3887}{\texttt{ arXiv:0802.3887}}.

\bibitem{CMS-PAS-EXO-11-030}
\hrefCMSnoop {} {{CMS Collaboration}, ``Search for third-generation leptoquarks
  and scalar bottom quarks in pp collisions at {$\sqrt{s}=7$~TeV}'',} (2012).
  \href{http://www.arXiv.org/abs/1210.5627}{\texttt{ arXiv:1210.5627}}.
  Submitted to JHEP.

\bibitem{lambdalimit}
\hrefCMSnoop {} {G.~Bhattacharyya, ``$R$-parity-violating supersymmetric Yukawa
  couplings: A mini-review'',} \textit{ Nucl. Phys. B Proc. Suppl.} \textbf{
  52} (1997) 83,
  \href{http://dx.doi.org/10.1016/S0920-5632(96)00539-7}{\doi{10.1016/S0920-5632(96)00539-7}},
  \href{http://www.arXiv.org/abs/hep-ph/9608415}{\texttt{
  arXiv:hep-ph/9608415}}.

\bibitem{CMSdetector}
\hrefCMSnoop {} {{ CMS} Collaboration, ``The {CMS} experiment at the {CERN}
  {LHC}'',} \textit{ JINST} \textbf{ 03} (2008) S08004,
\href{http://dx.doi.org/10.1088/1748-0221/3/08/S08004}{\doi{10.1088/1748-0221/3/08/S08004}}.

\bibitem{CMS-PAS-PFT-09-001}
\href {http://cdsweb.cern.ch/record/1194487} {{ CMS} Collaboration,
  ``Particle--Flow Event Reconstruction in {CMS} and Performance for Jets,
  Taus, and {\MET}'',} CMS Physics Analysis Summary CMS-PAS-PFT-09-001, (2009).

\bibitem{HPS}
\hrefCMSnoop {} {{ CMS} Collaboration, ``Performance of $\tau$-lepton
  reconstruction and identification in CMS'',} \textit{ JINST} \textbf{ 07}
  (2012) P01001,
  \href{http://dx.doi.org/10.1088/1748-0221/7/01/P01001}{\doi{10.1088/1748-0221/7/01/P01001}},
  \href{http://www.arXiv.org/abs/1109.6043}{\texttt{ arXiv:1109.6043}}.

\bibitem{Cacciari:2008gp}
\hrefCMSnoop {} {M.~Cacciari, G.~P. Salam, and G.~Soyez, ``{The anti-$k_T$ jet
  clustering algorithm}'',} \textit{ JHEP} \textbf{ 04} (2008) 063,
  \href{http://dx.doi.org/10.1088/1126-6708/2008/04/063}{\doi{10.1088/1126-6708/2008/04/063}},
  \href{http://www.arXiv.org/abs/0802.1189}{\texttt{ arXiv:0802.1189}}.

\bibitem{Cacciari:JetArea}
\hrefCMSnoop {} {M.~Cacciari and G.~P. Salam, ``{Pileup subtraction using jet
  areas}'',} \textit{ Phys. Lett. B} \textbf{ 659} (2008) 119,
  \href{http://dx.doi.org/10.1016/j.physletb.2007.09.077}{\doi{10.1016/j.physletb.2007.09.077}},
  \href{http://www.arXiv.org/abs/0707.1378}{\texttt{ arXiv:0707.1378}}.

\bibitem{BTV-11-004}
\href {https://cdsweb.cern.ch/record/1427247} {{ CMS} Collaboration, ``b-Jet
  identification in the CMS experiment'',} CMS Physics Analysis Summary
  CMS-PAS-BTV-11-004, (2011).

\bibitem{Sjostrand:2003wg}
\hrefCMSnoop {} {T.~Sj{\"o}strand, S.~Mrenna, and P.~Skands, ``PYTHIA 6.4
  physics and manual'',} \textit{ JHEP} \textbf{ 05} (2006) 026,
  \href{http://dx.doi.org/10.1088/1126-6708/2006/05/026}{\doi{10.1088/1126-6708/2006/05/026}},
  \href{http://www.arXiv.org/abs/hep-ph/0603175}{\texttt{
  arXiv:hep-ph/0603175}}.

\bibitem{Alwall:2011uj}
J.~Alwall\hrefCMSnoop {} { {et~al.}, ``{MadGraph 5: going beyond}'',} \textit{
  JHEP} \textbf{ 06} (2011) 128,
  \href{http://dx.doi.org/10.1007/JHEP06(2011)128}{\doi{10.1007/JHEP06(2011)128}},
  \href{http://www.arXiv.org/abs/1106.0522}{\texttt{ arXiv:1106.0522}}.

\bibitem{POWHEG2}
\hrefCMSnoop {} {{S. Frixione, P. Nason, C. Oleari}, ``Matching {NLO QCD}
  computations with parton shower simulations: the {POWHEG} method'',} \textit{
  JHEP} \textbf{ 11} (2007) 070,
  \href{http://dx.doi.org/10.1088/1126-6708/2007/11/070}{\doi{10.1088/1126-6708/2007/11/070}},
  \href{http://www.arXiv.org/abs/0709.2092}{\texttt{ arXiv:0709.2092}}.

\bibitem{TAUOLA}
\hrefCMSnoop {} {Z.~W\c{a}s, ``{TAUOLA} the library for tau lepton decay, and
  {KKMC/KORALB/KORALZ}\ldots status report'',} \textit{ Nucl. Phys. B, Proc.
  Suppl.} \textbf{ 98} (2001) 96,
  \href{http://dx.doi.org/10.1016/S0920-5632(01)01200-2}{\doi{10.1016/S0920-5632(01)01200-2}}.

\bibitem{Agostinelli2003250}
\hrefCMSnoop {} {S.~Agostinelli {et~al.}, ``Geant4---a simulation toolkit'',}
  \textit{ Nucl. Instrum. Meth. A} \textbf{ 506} (2003) 250,
  \href{http://dx.doi.org/10.1016/S0168-9002(03)01368-8}{\doi{10.1016/S0168-9002(03)01368-8}}.

\bibitem{LQVLQ}
A.~Belyaev\hrefCMSnoop {} { {et~al.}, ``Leptoquark Single and Pair production
  at LHC with CalcHEP/CompHEP in the complete model'',} \textit{ JHEP} \textbf{
  0509} (2005) 005,
  \href{http://dx.doi.org/10.1088/1126-6708/2005/09/005}{\doi{10.1088/1126-6708/2005/09/005}},
  \href{http://www.arXiv.org/abs/0502067}{\texttt{ arXiv:0502067}}.

\bibitem{CalcHEP}
\hrefCMSnoop {} {A.~Belyaev, N.~D. Christensen, and A.~Pukhov, ``CalcHEP 3.4
  for collider physics within and beyond the Standard Model'',} (2012).
  \href{http://www.arXiv.org/abs/1207.6082}{\texttt{ arXiv:1207.6082}}.

\bibitem{prospino}
W.~Beenakker\hrefCMSnoop {} { {et~al.}, ``Squark and gluino production at
  hadron colliders'',} \textit{ Nucl. Phys. B} \textbf{ 492} (1997) 51,
  \href{http://dx.doi.org/10.1016/S0550-3213(97)80027-2}{\doi{10.1016/S0550-3213(97)80027-2}}.

\bibitem{LQ-sigmaNLO}
M.~Kr{\"a}mer\hrefCMSnoop {} { {et~al.}, ``Pair production of scalar
  leptoquarks at the {CERN LHC}'',} \textit{ Phys. Rev. D} \textbf{ 71} (2005)
  057503,
  \href{http://dx.doi.org/10.1103/PhysRevD.71.057503}{\doi{10.1103/PhysRevD.71.057503}},
  \href{http://www.arXiv.org/abs/hep-ph/0411038}{\texttt{
  arXiv:hep-ph/0411038}}.

\bibitem{FEWZ}
\hrefCMSnoop {} {{K.~Melnikov and F.~Petriello}, ``Electroweak gauge boson
  production at hadron colliders through O($\alpha_{s}^2$)'',} \textit{ Phys.
  Rev. D} \textbf{ 74} (2006) 114017,
  \href{http://dx.doi.org/10.1103/PhysRevD.74.114017}{\doi{10.1103/PhysRevD.74.114017}},
  \href{http://www.arXiv.org/abs/hep-ph/0609070}{\texttt{
  arXiv:hep-ph/0609070}}.

\bibitem{Kidonakis:2010}
\hrefCMSnoop {} {N.~Kidonakis, ``Next-to-next-toleading soft-gluon corrections
  for the top quark cross section and transverse momentum distribution'',}
  \textit{ Phys. Rev. D} \textbf{ 82} (2010) 114030,
  \href{http://dx.doi.org/10.1103/PhysRevD.82.114030}{\doi{10.1103/PhysRevD.82.114030}},
\href{http://www.arXiv.org/abs/1009.4935}{\texttt{ arXiv:1009.4935}}.

\bibitem{CMS-PAS-SMP-12-008}
\href {http://cdsweb.cern.ch/record/1434360} {{CMS Collaboration}, ``Absolute
  Calibration of the Luminosity Measurement at CMS: Winter 2012 Update'',} CMS
  Physics Analysis Summary CMS-PAS-SMP-12-008, (2012).

\bibitem{CMS-EWK-DIW}
\hrefCMSnoop {} {{ CMS} Collaboration, ``Measurement of {${\rm W}^{+}{\rm
  W}^{-}$} production and search for the {H}iggs boson in pp collisions at
  $\sqrt{s}=7$~{TeV}'',} \textit{ Phys. Lett. B} \textbf{ 699} (2011) 25,
  \href{http://dx.doi.org/10.1016/j.physletb.2011.03.056}{\doi{10.1016/j.physletb.2011.03.056}}.

\bibitem{CMS-EWK-11-012}
\hrefCMSnoop {} {{ CMS} Collaboration, ``Measurement of the
  {Z$/\gamma^*$+\,b-jet} cross section in pp collisions at
  {$\sqrt{s}=7$\TeV}'',} \textit{ JHEP} \textbf{ 06} (2012) 126,
  \href{http://dx.doi.org/10.1007/JHEP06(2012)126}{\doi{10.1007/JHEP06(2012)126}}.

\bibitem{PDF4LHC1}
\hrefCMSnoop {} {M.~Botje {et~al.}, ``{The PDF4LHC Working Group Interim
  Recommendations}'',} (2011).
  \href{http://www.arXiv.org/abs/1101.0538}{\texttt{ arXiv:1101.0538}}.

\bibitem{LHC-HCG}
\href {http://cdsweb.cern.ch/record/1379837} {{ATLAS and CMS Collaborations,
  LHC Higgs Combination Group}, ``Procedure for the {LHC} {H}iggs boson search
  combination in {S}ummer 2011'',} CERN Report ATL-PHYS-PUB-2011-11 and CMS
  NOTE-2011/005, (2011).

\bibitem{Brooijmans:2012yi}
G.~Brooijmans\hrefCMSnoop {} { {et~al.}, ``{Les Houches 2011: Physics at TeV
  Colliders New Physics Working Group Report}'',} (2012).
\href{http://www.arXiv.org/abs/1203.1488}{\texttt{ arXiv:1203.1488}}.

\bibitem{springerlink:10.1007/JHEP08(2012)026}
\hrefCMSnoop {} {{ CMS} Collaboration, ``Search for stopped long-lived
  particles produced in pp collisions at $\sqrt{s}=7$~TeV'',} \textit{ JHEP}
  \textbf{ 08} (2012) 12,
  \href{http://dx.doi.org/10.1007/JHEP08(2012)026}{\doi{10.1007/JHEP08(2012)026}}.

\end{thebibliography}\endgroup
